\documentclass[%
 amsmath,amssymb,
 aps,
 longbibliography,
 twocolumn,
prb,
floatfix,
]{revtex4-2}

\usepackage{graphicx}
\usepackage{dcolumn}
\usepackage{bm}
\usepackage[colorlinks=true, citecolor=red, linkcolor=blue, urlcolor=black]{hyperref}

\usepackage{enumerate}
\usepackage{float}

\newcommand{\ket}[1]{\lvert#1\rangle}

\newcommand{\mean}[1]{\langle#1\rangle}

\newcommand{\K}{\operatornamewithlimits{\mathcal{K}}}

\usepackage[dvipsnames]{xcolor}
\usepackage[normalem]{ulem}
\usepackage{ifthen}

\newcounter{verbosity}
\begin{document}


\title{Rare collapse of fermionic quasiparticles 
upon coupling to local bosons
}

\author{Piotr Wrzosek}
\affiliation{%
Institute of Theoretical Physics, Faculty of Physics, University of Warsaw, Pasteura 5, PL-02093 Warsaw, Poland
}%
\author{Adam K\l{}osi\'n{}ski}
\affiliation{%
Institute of Theoretical Physics, Faculty of Physics, University of Warsaw, Pasteura 5, PL-02093 Warsaw, Poland
}%
\author{Krzysztof Wohlfeld}
\affiliation{%
Institute of Theoretical Physics, Faculty of Physics, University of Warsaw, Pasteura 5, PL-02093 Warsaw, Poland
}%
\author{Cli\`o{} Efthimia Agrapidis}
\email[Correspondence to: ]{clio.agrapidis@fuw.edu.pl}
\affiliation{%
Institute of Theoretical Physics, Faculty of Physics, University of Warsaw, Pasteura 5, PL-02093 Warsaw, Poland
}%

\date{\today}

\begin{abstract}
We study the stability of the fermionic quasiparticle in a fermion-boson model on a Bethe lattice, 
with fermions interacting with local bosons 
via {\color{black}polaronic-type coupling}. We solve the problem
by mapping it onto a non-interacting
chain with site-dependent potential. 
We show that, despite a finite number of bosonic excitations costing zero energy, {\color{black} among the many analyzed cases, the occurrence of a complete collapse of the quasiparticle is rare.}
The quasiparticle disappearance becomes easier with an increase in: (i) the total number of bosons with zero energy, and
(ii) the relative strength of the coupling between bosons and fermions. The postulated model can, among other things, be applied to study systems in which 
fermions are introduced into antiferromagnetic (or antiferro-orbital) domains surrounded by ferromagnetic (or ferro-orbital) ordered states. This might take place in the overdoped cuprates or upon doping manganese or vanadium oxides.
Finally, we show how this model leads to an in-depth understanding of the onset of quasiparticles in the 1D and 2D $t$-$J^z$ model.
\end{abstract}

\maketitle


\section{\label{sec:intro}Introduction}

One of the most standard approaches to tackle a quantum many-body system is to approximate it by  weakly-interacting long-lived quasiparticles~\cite{Venema2016}. 
This is a very successful picture as destroying
a quasiparticle turns out to be a far more complex task than naively expected.
One of the first physical system posited to exhibit quasiparticle decay was $^4$He: already in 1959 Pitaevskii suggested that the phononic quasiparticle was allowed to 
decay into a two-roton continuum~\cite{Pitaevskii1959, Glyde2018}.
Yet, it is nowadays believed that the quasiparticle does not enter the two-roton continuum in $^4$He~\cite{Glyde2018}
and instead the decay is exponentially avoided due to  
strong interactions~\cite{Verresen2019, Gaveau1995}.
A somewhat similar situation takes place in some non-collinearly
ordered magnets with small spin. Here, the decay of magnon quasiparticles is also not observed~\cite{Ito2017, Verresen2019}. It is then only 
for large spins and weaker magnon interactions that the magnon decay takes place and can be observed~\cite{Chernyshev2009, Oh2013}. Nevertheless, these decays do not seem to overdamp the magnons in realistic spin models~\cite{Zhitomirsky2013}.

To search for the total quasiparticle collapse, often referred to as non-Fermi liquid or `unparticle' physics~\cite{Phillips2006, Zaanen2019} and realised in the high-temperature superconducting cuprates (cf.~\cite{Chen2019, Wahlberg2021}), one has to go to more exotic models. These typically concern fermions
coupled to a gapless mobile boson with a very particular type 
of fermion-boson coupling~\cite{Watanabe2014}.
The latter occurs when the gapless bosons
are Goldstone modes and the coupling to fermions
does {\it not} vanish in the limit of low energy-momentum transfer~\cite{Kane1989, Watanabe2014}. However, as suggested in Ref.~\cite{Watanabe2014}, such a coupling is relatively rare in nature (its best realisation being a nematic Fermi liquid). Another, probably more common, route is to explore interactions between massless gauge bosons~\cite{Powell2020}
and fermions~\cite{Lee1992, Altshuler1994, Chakravarty1995, Lee2009} which can lead to non-Fermi liquid behavior in two dimensions (2D). Such couplings may play a vital role in quantum Hall systems~\cite{Halperin1993}, spin liquids~\cite{Lee2005} or quantum critical systems such as heavy fermions~\cite{Gegenwart2008}. 

In this paper we explore yet another route to quasiparticle extinction. To this end, we adhere to a fermion-boson model with a {\color{black}bond-type polaronic coupling} between the two particle species. The primary difference with respect to all of the cases above is that the bosons in our model are immobile, i.e., local,
and the lattice translational symmetry is explicitely broken.
On the other hand, 
 also for the here-studied model,
it is important that a finite number of bosons can become massless (cost zero energy). While such a model is interesting {\it per se}, we believe it to be of relevance to, for instance, systems in which fermions 
 are introduced into antiferromagnetic (or antiferro-orbital) domains surrounded by ferromagnetic  (or ferro-orbital) ordered states. This might take place in overdoped cuprates \cite{Kopp2007, Jia2014, Santoso2017, Ong2022, Lee2006, Zhang2022, Kopp2007, Battisti2016} or upon doping manganese or vanadium oxides  \cite{Dagotto20011, Miyasaka2000, Fujioka2005, Fujioka2008, Avella2019}.

All of the results obtained in the paper follow from exact analytical diagonalisation of the Hamiltonian. This enables us to address the issue 
of quasiparticle stability in an unbiased way. The main result is that a complete quasiparticle collapse is in general possible in the fermion-boson model -- though only in 
a small portion of the model parameter space, making it a rare occurence.
The quasiparticle disappearance 
is allowed once a certain number of local bosonic excitations cost zero energy.
This decay becomes easier with an increase in: (i) the total number of bosons with zero energy, and
(ii) the relative strength of the coupling between bosons and fermions. 

Finally, we uncover an interesting relation between the model introduced in this paper and the well-studied problem of a mobile hole introduced in the one-dimensional (1D) or 2D Ising antiferromagnet -- as given by the 1D or 2D $t$--$J^z$ model~\cite{Bulaevskii1968, Brinkman1970, Kane1989, 
Starykh1996,  Sorella1998, 
Chernyshev1999,
Smakov2007,
Smakov2007b,
Maka2014,
Grusdt2018, Bieniasz2019, Wrzosek2021}.
It turns out that these models, which always support a quasiparticle solution, are a specific realisation of the class of fermion-boson models considered in this paper. This means that the hole in the Ising
antiferromagnet always forming a quasiparticle solution is not the generic one.  

The paper is organised as follows: in Sec. \ref{sec:ModelMethods} we introduce the fermion-boson Hamiltonian and the methods used to perform the calculations, including mapping the interacting model to a non-interacting chain with the same spectral properties; in Sec. \ref{sec:PointPotential} we study the effect of the coupling to the impurity-like bosons, i.e. bosons which all cost zero energy except for one particular site, on the appearance of a quasiparticle; then in Sec. \ref{sec:StringLike} we extend our study to string-like local bosons, i.e. bosons which have finite energy for a whole range of sites. Within Secs. \ref{sec:PointPotential} and \ref{sec:StringLike}, we also show how our results relate to the $t$-$J^z$ model on a 1D and 2D Bethe lattice, respectively. Lastly, we discuss the results and draw our conclusions in Sec. \ref{sec:Discussion}.





\section{Models}\label{sec:ModelMethods}

\subsection{Fermion-boson model}

We consider an interacting Hamiltonian on a Bethe lattice with coordination number $z$,
\begin{equation} \label{model}
\begin{aligned}
\mathcal{H} = - t &\sum_{\mean{i,j}} \left[ h_i^\dag h_j \left(a_i + a_j^\dag  \right)\right.+ \left. h_j^\dag h_i \left(a_j + a_i^\dag  \right)\right] \\ &+ \sum_i J_i a_i^\dag a_i,
\end{aligned}
\end{equation}
where $h_i$ are fermion annihilation operators and $a_i$ are hard-core bosons annihilation operators at site $i$, $t$ is the 
{\color{black} polaronic-type coupling leading to creation or annihilation of bosons once fermions hop across a bond (a bond-type polaronic coupling),} 
and $J_i$ is the on-site boson potential.  The model lives in a restricted Hilbert space including states with the constraint \begin{align}\label{constraint}
n_{a_i}+n_{h_i} \leq 1, 
\end{align}
with $n_{a_i}$ and $n_{h_i}$ being the number operator at site $i$ for bosons and fermions respectively.

{\color{black}
We note that constraint \eqref{constraint}
is motivated here by the discussed later in the text connections between this model and 
{\it various} variants of the $t$--$J^z$ models -- see Sec.~\ref{sec:pointphysreal} and 
Sec.~\ref{sec:physrealstringlike}. The use of such a constraint allows for reducing the fermion-boson Hilbert space to the one spanned by the eigenstates of the $t$--$J^z$ models, see \eqref{eq:mappings}. On the other hand, such a constraint means that the considered fermion-boson model is rather not a realistic description of an electron-phonon problem -- albeit a comparison against the latter one is envisaged in the future.
}

\begin{figure}[t]
    \centering
    \includegraphics{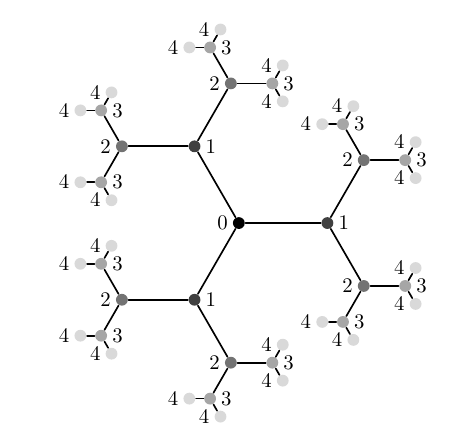}
    \caption{Example of the distance labelling on a Bethe lattice with $z=3$. The numbering corresponds to the distance $d_i$ from the root of the Bethe lattice, labelled as $0$.}
    \label{fig:distances}
\end{figure}

We chose one arbitrary site for the root of the Bethe lattice. We label this site with index 0 and we will call it an origin of the lattice. Let $d_i$ stand for the distance of site $i$ to the origin of the lattice, calculated in the number of edges in the graph of the lattice that separate the site i from site 0, as shown in Fig.~\ref{fig:distances}. With this, we impose a geometrical constraint on the shape of the on-site boson potentials $J_i$ that renders equivalent all the branches of the Bethe lattice (starting at site 0),
\begin{equation}
d_i = d_j \Rightarrow J_i = J_j.
\end{equation}

In this work we are interested in the single-fermion spectral function 
\begin{equation} \label{spectral-function}
\mathcal A(\omega) = -\frac{1}{\pi}\text{Im} \mathcal{G}(\omega + i\delta),
\end{equation}
where the local single-particle Green's function $\mathcal G$  is given by:
\begin{equation}\label{GreensFunctionBethe}
    \mathcal G(\omega + i\delta) = \left\langle \varnothing \left| h_0 \frac{1}{\omega - \mathcal{H}
    + i\delta
    } h^\dag_0 \right| \varnothing
    \right\rangle,
\end{equation}
where $\ket{\varnothing}$ denotes the vacuum state for the particles in~Eq.~\eqref{model}
and the fermion $h$ is created at site $0$ of the Bethe lattice. Here the local Green's function describes the motion of a single fermion in the considered Hamiltonian \eqref{model}, for it describes the case when a single fermion is added at the site labelled as 0 of the Bethe lattice, then propagates via Hamiltonian (\ref{modelham}), and finally it is annihilated at site 0.
Note that we choose to
probe the system using the {\it local} Green's function,
for the model we consider is anisotropic in real space
(through, among other features, the site-dependent boson energies $J_i$).

\subsection{Mapping to a non-interacting model}

Since we are interested in the spectral function \eqref{spectral-function}, we can restrict ourselves to the subspace of states that contribute to it [see below]. Within this subspace, the above model in Eq.~\eqref{model} can be mapped to a single spinless fermion moving on an otherwise empty 1D chain with nearest-neighbour hopping and an external potential. The hopping amplitudes are equal everywhere other than between the 0-th site at the center of the chain and its two neighbors and the chain is placed in a site-dependent external potential that is symmetric around the 0-th site. The mapping is done via

\begin{equation}
V_{i} = \sum_{d_n=0}^{i-1} J_n \quad
{\rm and}\quad
\left\{ \begin{array}{c}
    \tau_0 = t \sqrt{z}\\
    \tau = t \sqrt{z-1}
    \end{array} \right. 
    \label{tjz-parametrization}
\end{equation}
and the non-interacting Hamiltonian is given by 
\begin{equation} \label{modelham}
    \begin{aligned}
     H &= \sum^\infty_{i=-\infty} \; \left[ -\tau \left( c^\dag_{i+1} c_i + \text{h.c.} \right) +  \; V_{|i|} \; n_i \right] \\
    &+\left( \tau - \frac{\tau_0}{\sqrt{2}} \right) \left( c^\dag_{1} c_0 + c_0^\dag c_{-1} + \text{h.c.} \right),
    \end{aligned} 
\end{equation}
where $c^\dag_i (c_i)$ creates (annihilates) a spinless fermion at site $i$, $n_i= c^\dag_i c_i$ is the fermion density operator at site $i$, $\tau$ and $\tau_0$ are the hopping amplitudes and $V_{|i|}$ is an external potential taken to be symmetric around the origin $i=0$, where the fermion is originally introduced in the system. {\color{black} We note that Eq.~\eqref{modelham} is equivalent to Eq.~\eqref{model} only thanks to some specific properties of the latter model: First,  we look at the spectral function of a single hole calculated for the hole introduced into the vacuum state of both holes and bosons. Second, there are no fluctuations of the bosonic degrees of freedom that are independent from the motion of the hole. And third, we consider either a chain or a Bethe lattice geometry. Altogether, the position of the bosons in our system is completely determined by the position of the hole. Thus, the bosonic part of the Hilbert space can be projected out and effectively included (in an exact manner) in the external potential term.  Hence, the mapping between the spectral properties, which are the primary interest of our study, holds.}
Without loss of generality we assume that the hopping amplitude $\tau> 0$.

Since we are interested in the single-fermion spectral function \eqref{spectral-function},
we need to define the relevant quantity also for the non-interacting 
model:
\begin{align}
A(\omega) = -\frac{1}{\pi}\text{Im} {G}(\omega + i\delta).
\label{eq:SpectralFunction}
\end{align}
where the corresponding single-particle Green's function is 
\begin{equation}
     G(\omega + i\delta) = \left\langle \varnothing \left| c_0 \frac{1}{\omega - {H} + i\delta } c^\dag_0 \right| \varnothing \right\rangle.
\end{equation}
Now, $\ket{\varnothing}$ denotes the vacuum for $c$ fermions, i.e. the empty 1D chain. 

Model (\ref{modelham}-\ref{eq:SpectralFunction}) is an effective model resulting from the mapping of model (\ref{model}-\ref{spectral-function}), but it can also be seen as a model describing a fermion in a 1D crystal (or optical lattice) with a particular pattern of impurities given rise to the potential $V_{|i|}$, cf. \cite{Moghaddam2021} for a recent work on a related problem.

\subsection{Solution} \label{Sec:GeneralSolution}

Our goal is to calculate the spectral function defined in Eq.~\eqref{eq:SpectralFunction}. To this end we choose the basis $\mathcal{B}$ in the following manner:
We start with the initial state $\ket{0} \equiv c^\dag_0 \ket{\varnothing}$, which corresponds to a spinless fermion located at site $i=0$. From this state we construct states `reachable' by the propagator $1/(\omega - {H} + i\delta)$ by repeated application of the Hamiltonian on our initial state, namely $H^n \ket{0}$, for some $n \in \mathbb{N}$. This leads to basis $\mathcal{B}$ defined as
\begin{equation}
\begin{split}
    \ket{0} &= c^\dag_0 \ket{\varnothing}, \\
    \ket{i} &= \frac{1}{\sqrt{2}} \left( c^\dag_{-i} + c^\dag_i \right) \ket{\varnothing}.
\end{split}
\end{equation}
In this basis, the Hamiltonian \eqref{modelham} takes the form $H=\sum_{i,j} h_{i,j} | i\rangle \langle j |$. Note that we can restrict the Hilbert subspace to $\mathcal{B}$, since all neglected  states, i.e. the anti-symmetric ones $\ket{i}_{-} = \frac{1}{\sqrt{2}} \left( c^\dag_{-i} - c^\dag_i \right) \ket{\varnothing}$, give zero contribution to the spectral function \cite{Wohlfeld2008}.

An essential feature of the introduced basis $\mathcal{B}$ is that the Hamiltonian matrix $[h_{i,j}]$  
becomes tridiagonal,
\begin{equation} \label{matrixham}
[h_{i,j}] = \begin{bmatrix}
V_0 & -\tau_0 &  &  &  \\
-\tau_0 & V_1 & -\tau &  &  \\
 & -\tau & V_2 & -\tau &  \\
 &  &  -\tau & V_3 & \ddots \\
 &  &  & \ddots & \ddots \\
\end{bmatrix}.
\end{equation}
This yields a simple formula for the Green's function in the form of a continued fraction,
\begin{equation}
 G(\omega) = \frac{1}{\omega - V_0 - \Sigma(\omega)}.
\label{GreensFunction}
\end{equation} 
Let us denote 
\begin{align}\label{TailExpansion}
    \Gamma(\omega) &= \frac{\tau^2}{\omega - V_l - \Gamma(\omega)}\\
    \Omega_{i<l} = & \omega - V_{i} \quad \text{and} \quad \Omega_l = \frac{\tau^2}{\Gamma(\omega)},
\end{align}
where $i\geq0$ denotes the distance from the origin. Then, we can write  the expression for the self-energy for a generic symmetric potential centered around $i=0$:
\begin{equation}
\begin{aligned}
\Sigma(\omega) = \frac{\tau_0^2}{\Omega_1 - \frac{\tau^2}{\Omega_2 - \frac{\tau^2}{\Omega_3 - \hdots}}} = \frac{\tau_0^2}{\tau^2}\K_{i=1}^l \left(\frac{\tau^2}{\Omega_{i}}\right).
\end{aligned}
\label{self-energy}
\end{equation}
{\color{black}
Note that in the considered shapes of $V_i$
{\it either} a finite number of terms in the continued fraction have to be calculated {\it or} an analytical solution to the continued fraction problem can
be achieved \cite{Bieniasz2019, Wrzosek2021}, see  Appendix \ref{App:A} for details. Therefore, all the obtained results can be considered to be numerically exact on infinite lattices.
}

In what follows, we are  interested in the properties of the state contributing to the spectral function $A(\omega )$ at the lowest energy (denoted as $\omega_{QP}$). In particular,
the central question 
is whether this state is a discrete one (i.e. a bound state) or is part of a continuum. As in the fermion-boson model this question corresponds to the issue of quasiparticle stability, we denote the spectral weight carried by the discrete state at $\omega_{QP}$
with the quasiparticle spectral weight $a_\mathrm{QP}$. In terms of the quasiparticle energy $\omega_{QP}$, this is given by:
\begin{equation} 
    \begin{aligned}
a_\mathrm{QP}&(\tau,\tau_0,V) = \lim_{\omega\to\omega_\mathrm{QP}} \left( \omega - \omega_\mathrm{QP} \right) \; G(\omega) = \\
&= \lim_{\omega\to\omega_\mathrm{QP}}\frac{1}{1-\frac{d}{d\omega}\Sigma(\omega)}
    \end{aligned}
    \label{QPweight}
\end{equation}
Hence, to obtain the expression for the quasi-particle residue we calculate the derivative of the self-energy with respect to the frequency $\omega$,
\begin{equation}
\begin{aligned}
\frac{d}{d\omega} \Sigma(\omega) &= \frac{\tau_0^2}{\tau^2} \frac{d}{d\omega} \K_{i=1}^n \left(\frac{\tau^2}{\Omega_{i}}\right) = \\
&= \frac{\tau_0^2}{\tau^4}\sum_{j=1}^n \left(-1 \right)^{j+1} \prod_{i=1}^j \left[ \K_{l=i}^n \left(\frac{\tau^2}{\Omega_l}\right) \right]^2 \frac{d\Omega_j}{d\omega},
\end{aligned}
\end{equation}
where $\frac{d}{d\omega}\Omega_{i<n} = 1$ and $\frac{d}{d\omega}\Omega_n = \frac{d}{d\omega}\left( \frac{\tau^2}{\Gamma(\omega)} \right)$.

Note that $\omega_\mathrm{QP}$ in \eqref{QPweight} depends on the specific form of the potential $V_{|i|}$, so we cannot provide a general form for this quantity at this point.

\section{Impurity-like bosons} \label{sec:PointPotential}


We consider a class of fermion-boson models with impurity-like bosons. To this end, we assume that the site-dependent bosonic energy $J_n$ in Eq.~\eqref{model} takes the form
\begin{equation}
    J_n=\begin{cases}
        J/2 & \mbox{if } d_n=0 \\
        0 & \mbox{if } d_n\neq 0
    \end{cases},
    \label{boson-poin-potential}
\end{equation}
i.e., all but one boson in the Bethe lattice are massless. This may look as a quite unphysical regime -- however, as shown below, it may find its realisation in a number of physical systems.

The above case corresponds to considering a point potential at site $n=0$ (i.e. at the position where we introduce the spinless fermion in the system) \eqref{modelham}
\begin{equation}
\label{eq:v}
    V_{n}=\begin{cases}
        0 & \mbox{if } n=0 \\
        V & \mbox{if } n\neq 0
    \end{cases},
\end{equation}
in the non-interacting model given in Eq.~\eqref{modelham}.

The other parameters of the two Hamiltonians, namely the coordination number $z$ in \eqref{model} and, correspondingly, the $\tau_0/\tau$ ratio in \eqref{modelham} are left free.
Since the two models are equivalent, we present the solution in terms of the easier-to-solve non-interacting model~\eqref{modelham}.

\subsection{\bf Solution}

\begin{figure}[t!]
\includegraphics[width=0.9\columnwidth]{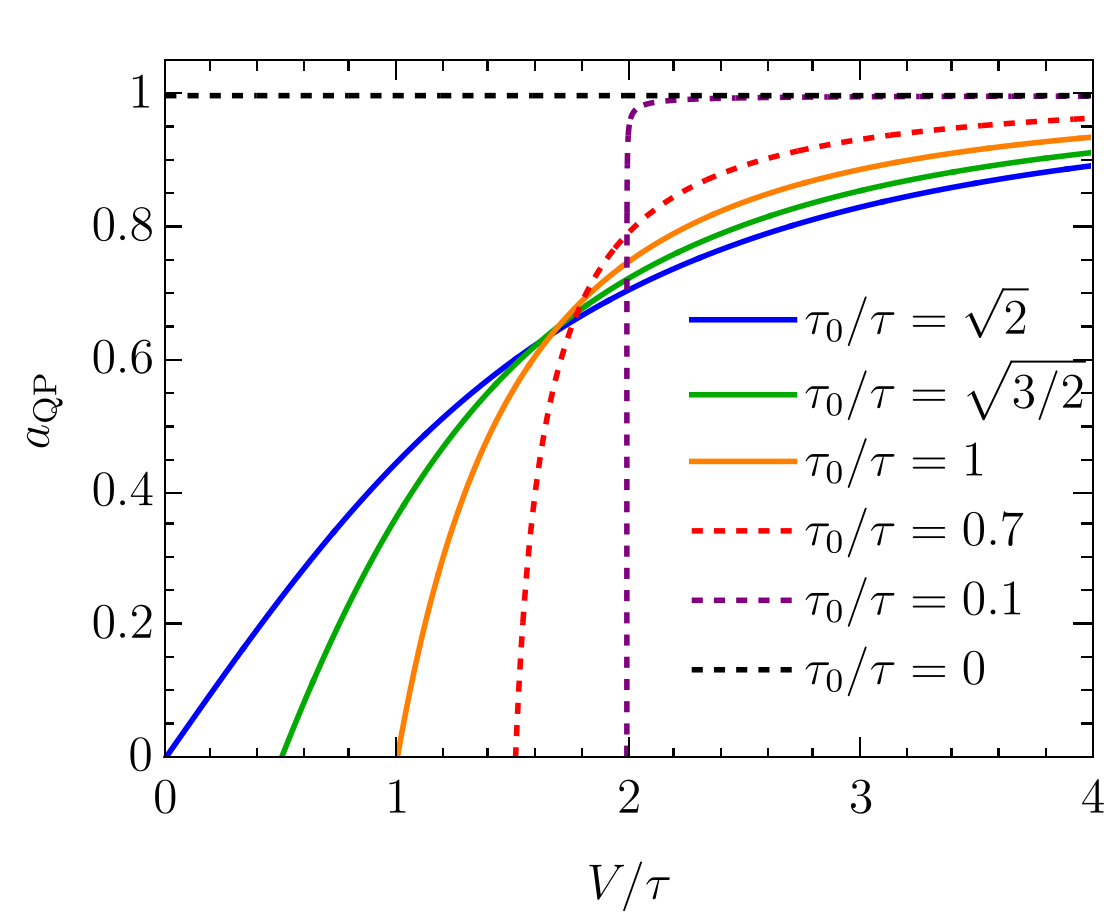}
\caption{The quasiparticle spectral weight $a_\mathrm{QP}(V)$ for  different values of ${\tau_0}/{\tau}$ as a function of the depth of the point potential $V$ [cf. Eq.~\eqref{eq:v}]. Note that the lines start at $V>0$ for any ${\tau_0}/{\tau}<\sqrt{2}$. Dashed lines shows values of $\tau_0/\tau$ for which the mapping to the Bethe lattice coordination number $z$ leads to a fractional value of the latter.}
\label{qpresidues}
\end{figure}


In what follows we leave the ratio  $\tau_0/\tau$ as a parameter in the system. However, 
we assume that $\tau_0 \leq \sqrt{2} \tau$. This is due to an earlier study~\cite{Wohlfeld2008}
which shows that larger values of $\tau_0$ always stabilise a quasiparticle
solution. In fact, $\tau_0 / \tau=\sqrt{2}$ is a limiting case which corresponds to a non-interacting isotropic chain [see Eq.~\eqref{modelham}] and, as shown below, is the case for which the quasiparticle solution
is always present, independently of the value of the point potential. 

\begin{figure}[t!]
\includegraphics[width=\columnwidth]{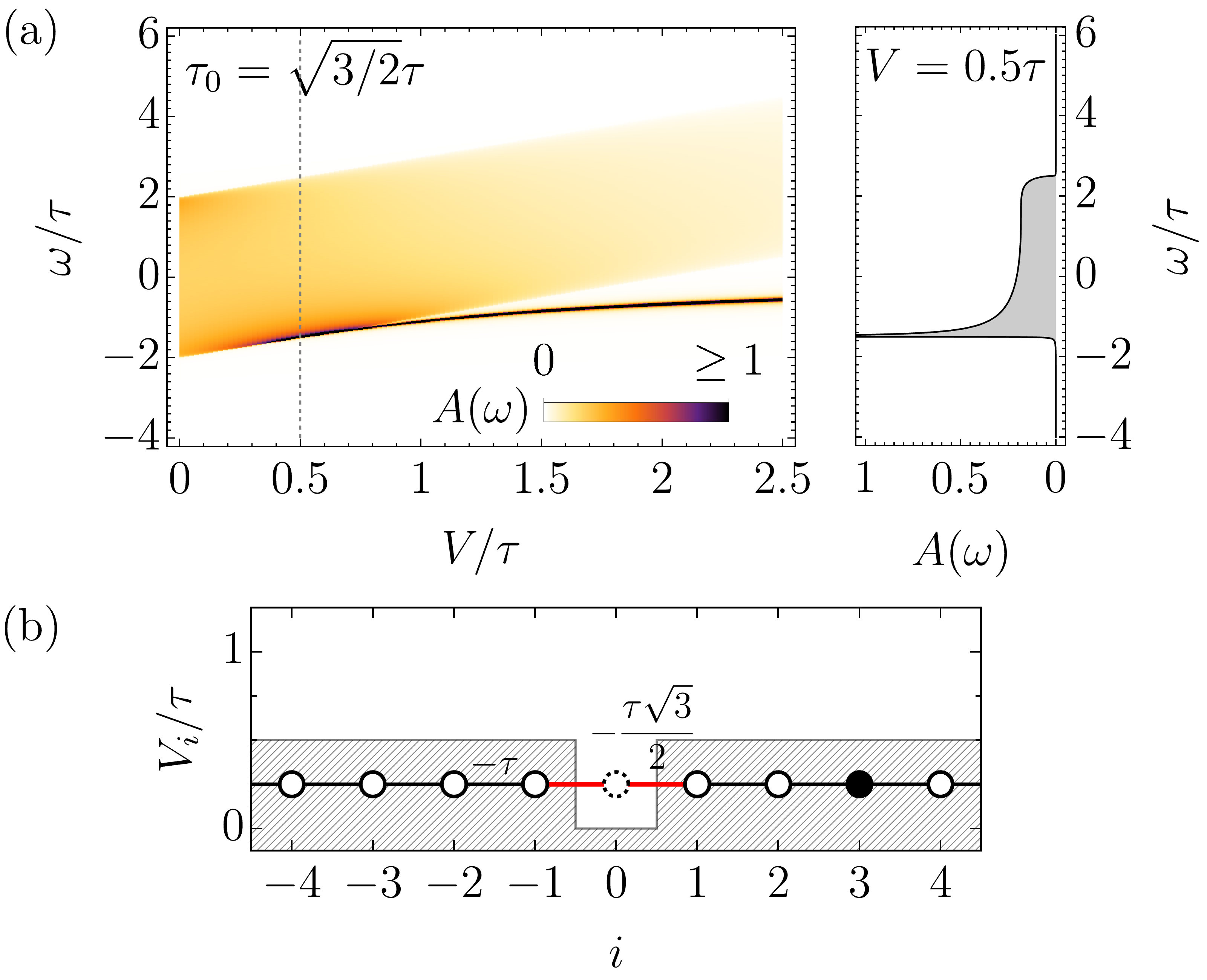}
\caption{(a) Spectral function $ A(\omega)$ as a function of the point-potential depth $V/\tau$ [cf. Eq.~\eqref{eq:v}] with ${\tau_0}/{\tau} = \sqrt{3/2}$. The right panel shows
the spectral function $A(\omega)$ at
$V=0.5\tau$: for this value no discrete energy peak is visible, since for
$\tau_0 / \tau = \sqrt{3/2}$ this corresponds to the critical value $V^*$ and
the quasiparticle appears only for $V>0.5\tau$.
(b) Schematic representation of the model at the vertical cut shown in (a) with ${\tau_0}/{\tau} = \sqrt{3/2}$ and $V=0.5\tau$. The different hopping around the origin is drawn in red, the gray area represents the shape of the potential, the black filled circle represents the moving fermion and the dashed circle its starting position. }
\label{pointpotential}
\end{figure}

The above assumption leads to the following expression for the self-energy \eqref{self-energy},

\begin{equation}
\Sigma(\omega) = \left\{ \begin{array}{lr}
\frac{\tau_0^2}{\tau^2} \left( \frac{\omega - V}{2} + \sqrt{ \left( \frac{\omega - V}{2} \right) ^2 - \tau^2} \right), & \: \omega \leq V\\
\frac{\tau_0^2}{\tau^2} \left( \frac{\omega - V}{2} - \sqrt{ \left( \frac{\omega - V}{2} \right) ^2 - \tau^2} \right), & \: \omega > V\\
\end{array} \right. .
\label{self-en-point}
\end{equation}
%
Moreover, following the general solution presented above in Sec. \ref{Sec:GeneralSolution}, we calculate the analytic formula for the quasiparticle spectral weight $a_\mathrm{QP}$ for the potential given in Eq.~\eqref{eq:v}. In terms of the quasiparticle energy $\omega_\mathrm{QP}$ we have
%
\begin{equation}
\begin{aligned}
a_\mathrm{QP}&(\tau,\tau_0,V) = \lim_{\omega\to\omega_\mathrm{QP}} \left( \omega - \omega_\mathrm{QP} \right) \; G(\omega) = \\
&= \lim_{\omega\to\omega_\mathrm{QP}}\frac{1}{1-\frac{d}{d\omega}\Sigma(\omega)} =\\
&= \left[ 1-\frac{1}{2} \frac{\tau_0^2}{\tau^2} \left( 1 + \frac{\omega_{\mathrm{QP}} - V}{\sqrt{(\omega_\mathrm{QP} - V)^2-4 \tau^2}} \right) \right]^{-1},
\end{aligned}
\label{QPresidue}
\end{equation}
where 
\begin{equation} \label{qppos}
    \omega_\mathrm{QP}= \left\{ 
    \begin{array}{lr}
    \frac{\tau_0^2}{2} \left(\frac{V}{\tau_0^2-\tau^2} - \frac{\sqrt{V^2 + 4\tau_0^2 - 4\tau^2}}{|\tau_0^2-\tau^2|} \right) & \tau_0 > \tau, 0\\
    - \frac{1}{V} & \tau_0 = \tau > 0\\
    \frac{\tau_0^2}{2} \left(\frac{V}{\tau_0^2-\tau^2} + \frac{\sqrt{V^2 + 4\tau_0^2 - 4\tau^2}}{|\tau_0^2-\tau^2|} \right) & 0 < \tau_0 < \tau\\
    0 & \tau_0 = 0
    \end{array} \right. .
\end{equation}
All of the above solutions come from the $\omega \leq V$ branch in (\ref{self-en-point}). We plot the quasiparticle weight $a_\mathrm{QP}$ as a function of $V/\tau$ for different values of $\tau_0/\tau$ in Fig.~\ref{qpresidues} and notice that the value of the potential $V$ at which the quasiparticle weight is finite shifts away from zero with tuning $\tau_0/\tau$ away from $\sqrt 2$.
We then determine for which depth of the potential $V$  the quasiparticle weight $a_\mathrm{QP}$ becomes finite as the hopping ratio $\tau_0/\tau$ changes. This leads to the definition of the critical value of $V^*$ 
\begin{equation} \label{critical-point-potential}
V^*(\tau,\tau_0) = \left\{
\begin{array}{lr}
\frac{2 \tau^2 - \tau_0^2}{|\tau|} & \tau_0 > 0\\
- \infty & \tau_0 = 0\\
\end{array} \right. .
\end{equation}
%
We can then also calculate the spectral function \eqref{eq:SpectralFunction} by using the self-energy in Eq. \eqref{self-en-point} and substituting in Eq. \eqref{GreensFunction}. We show an example of the spectral function for the model \eqref{modelham}-\eqref{eq:v} with $\tau_0/\tau=\sqrt{3/2}$ in Fig.~\ref{pointpotential}. From Eq. \eqref{critical-point-potential}, we find $V^*=0.5\tau$. Indeed, no discrete energy branch is visible for $V\le V^*$, $V^*$ being the critical value of the point potential for which the discrete state appears.

 Let us now go back to the fermion-boson interacting Hamiltonian \eqref{model}. We are considering a boson potential as given by \eqref{boson-poin-potential}, while we can vary the Bethe lattice coordination number $z$. Substituting \eqref{tjz-parametrization} in \eqref{critical-point-potential}, we find the critical value of the boson potential $J^*$
\begin{equation}
    J^*(t,z)=
        \frac{t (z-2)}{\sqrt{z-1}}.
    \label{J-critical}
\end{equation}
Note that $z$  can only take integer values equal or larger than two, so that we lose part of the solution available when considering the non-interacting chain \eqref{modelham}. Condition \eqref{J-critical} implies that, also for the interacting fermion-boson model on a Bethe lattice with coordination number $z$ \eqref{model}, there are two distinct regimes: one with a well defined fermionic quasiparticle for $J> J^*$, and one with no 
quasiparticle for $J\leq J^*$.

\subsection{Special limit}

\begin{figure}[t!] 
    \includegraphics[width=\columnwidth]{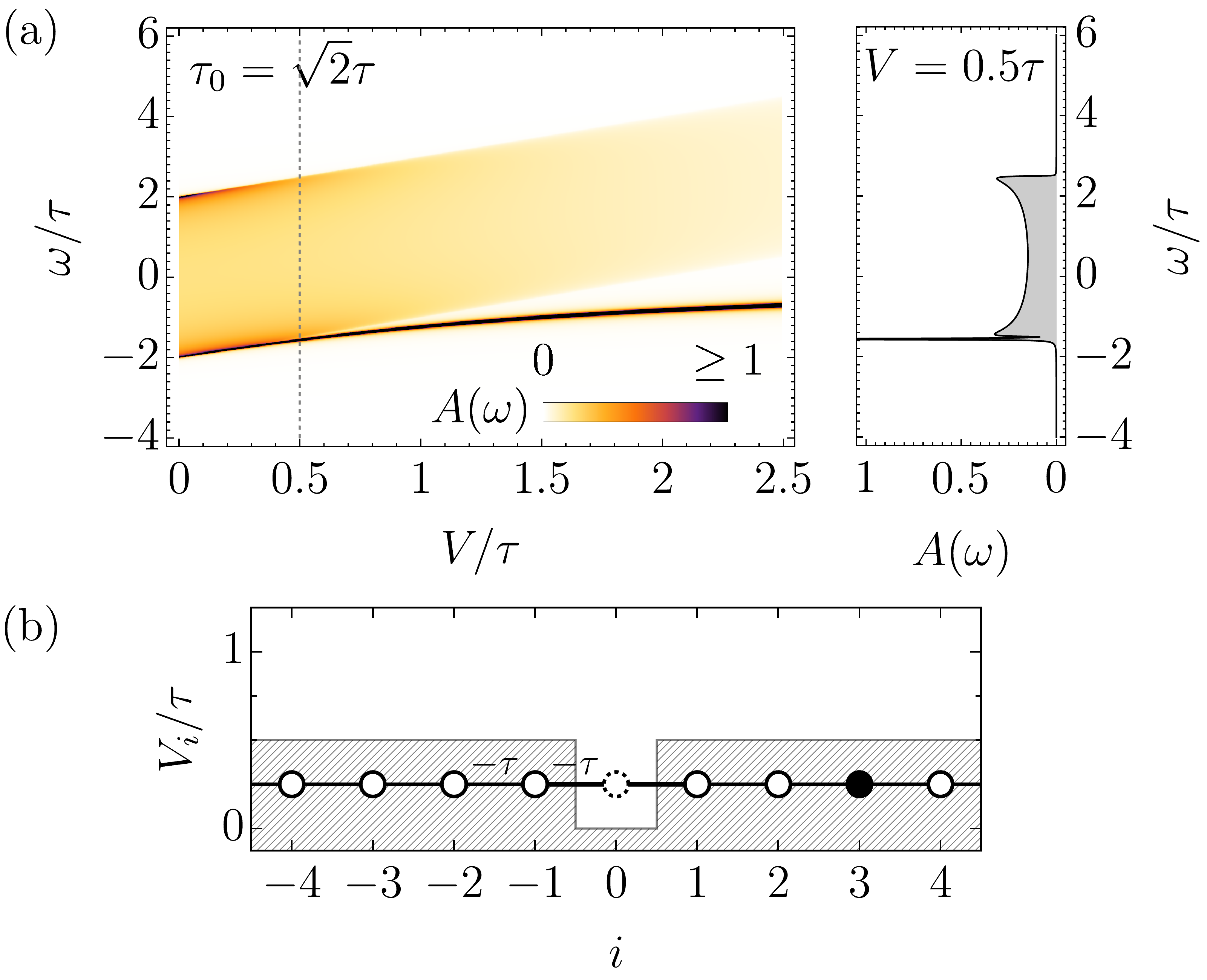}
    \caption{Spectral function $A(\omega)$ as a function of the potential depth $V/\tau$ [cf. Eq.~\eqref{eq:v}]  for the `special limit' $\tau_0/\tau=\sqrt{2}$. The right panel shows the spectral function $A(\omega)$ at $V=0.5\tau$: for this value there is a clear discrete energy state split from a continuum, since for $\tau_0 / \tau = \sqrt{2}$
    we have the critical
    $V^* =0$
    and the quasiparticle appears for any $V>0$.
    (b) Schematic representation of the model at the vertical cut shown in (a) with ${\tau_0}/{\tau} = \sqrt{2}$ and $V=0.5\tau$. In this case, all of the hoppings in the non-interacting chain are equivalent. The gray area represents the shape of the potential, the black filled circle represents the moving fermion and the dashed circle its starting position.}
    \label{1Dcase}
\end{figure}

Fig.~\ref{qpresidues} shows the existence of a special value $\tau_0/\tau=\sqrt{2}$ for which the discrete state exists for any value $V>0$. We then consider the spectral function $A(\omega)$ for this case (see Fig.~\ref{1Dcase}). Indeed, a discrete energy branch is present for all values $V>0$. As shown in Fig.~\ref{qpresidues}(b), taking $\tau_0/\tau=\sqrt{2}$ corresponds to considering a chain with equal hoppings on all bonds, i.e., the hopping from the origin is not distinct from the others anymore. 
In terms of the fermion-boson model \eqref{model}, this translates to the existence of a special value of the coordination number $z$ at which the quasiparticle is present for any value $J>0$: $z=2$ (corresponding to $\tau_0/\tau=\sqrt 2$). This corresponds to a 1D interacting fermion-boson model with a boson point potential. The latter system is equivalent to a well-known interacting model: the $t$-$J^z$ chain with one single hole (see below).

\subsection{Physical realization}
\label{sec:pointphysreal}

\begin{figure}[t!]
    \centering
    \includegraphics[width=0.8\columnwidth]{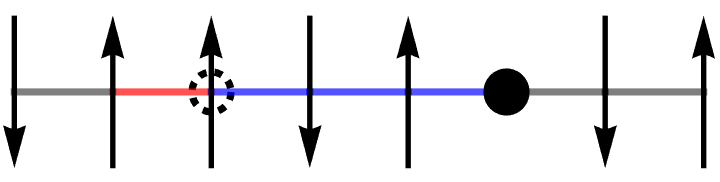}
    \caption{Dynamics of a single hole on a 1D chain $t$-$J^z$ (Bethe lattice with $z=2$). The black solid circle represents the hole in the current state, the dashed circle is the hole at its initial creation site. The cartoon shows the system after few hole hoppings have occured.}
    \label{fig:1DtJz}
\end{figure}

{\it $z=2$ case}
We have seen how a critical value of the relative strength of the fermion-boson coupling $J^*/t$ determines the existence or suppression of a quasiparticle and we have determined a critical case for the coordination number $z=2$ at which the quasiparticle exists for any $J/t>0$.
For this special case, we now show how we recover the physics of one of the most studied interacting models in condensed matter physics: the $t$-$J^z$ chain with one single hole. Indeed, we will show how its spectral function and of the fermion-boson model \eqref{model} (with $z=2$) are the same.

The Hamiltonian of the 1D $t\text{-}J^z$~model with a single hole reads
\begin{equation} \label{tjz}
\mathcal{H}_{t\text{-}J^z} = -t \sum_{\mean{i,j},\sigma} \left(
\tilde{c}_{i\sigma}^{\dag}\tilde{c}_{j\sigma}^{}\!+\!\mathrm{H.c.}\!\right)
+ J^z \sum_{\mean{i,j}}\left( S_{i}^{z}S_{j}^{z}\!
-\tfrac14\,\tilde{n}_i\tilde{n}_j\right),
\end{equation}
where $\tilde{c}_{i\sigma}$ annihilates an electron at site $i$ in the constrained Hilbert space without double occupancy, $t$ is the nearest neighbor hopping parameter, $S^z_i$ is the $z$-component of the spin-1/2 operator at site $i$, $\tilde n_i=\tilde{c}_{i\sigma}^\dagger \tilde{c}_{i\sigma}$ is the electron density operator at site $i$ and $J^z$ is the antiferromagnetic exchange coupling.
%
%
Rather than using the electron and spin operators, we can rewrite the Hamiltonian \eqref{tjz} in terms of hole and magnon operators \cite{Holstein1940}. The transformation starts with the rotation of all spins on one sublattice,
which in turn allows for the introduction of holes and magnons in terms of the following transformations:

\begin{equation} \label{eq:mappings}
\begin{aligned}
\tilde{c}_{i\uparrow}^\dag &= h_i P_i, &&  \tilde{c}_{i\uparrow} = P_i h_i^\dag , \\
\tilde{c}_{i\downarrow}^\dag &= h_i a_i^\dag, && \tilde{c}_{i\downarrow} = h_i^\dag a_i,\\
 S_i^z &= \frac{1}{2} - a_i^\dag a_i - \frac{1}{2}h_i^\dag h_i, &&
 \tilde{n}_i = 1 - h_i^\dag h_i,\\
 P_i&=\sqrt{1 - a_i^\dag a_i}
\end{aligned}
\end{equation}
where $h_i$ is a fermion operators that annihilates a hole and $a_i$ is a bosonic operator annihilating a magnon at site $i$. 
The kinetic energy now reads:

\begin{equation}\label{eq:kinetic}
\begin{aligned}
\mathcal{H}_t = \mathcal P \Big\{ -t \sum_{\mean{i,j}}  &\left[ h_i^\dag h_j \left( a_i + a_j^\dag  \right)  +  h_j^\dag h_i \left(  a_j + a_i^\dag \right) \right] 
\Big\} \mathcal P,
\end{aligned}
\end{equation}
with $\mathcal P$ being the global action of the projection operators $P_i$ and acting by projecting out states that do not satisfy $n_{a_i}+n_{h_i}\leq 1$. The potential energy reads

\begin{equation} \label{hammagint}
\begin{aligned}
\mathcal{H}_{J^z} &= E_0 + \frac{J^z}{2} \sum_{\mean{i,j}} \left[ a_i^\dag a_i + a_j^\dag a_j - 2 a_i^\dag a_i a_j^\dag a_j \right. + \\
&+\left. h_i^\dag h_i + h_j^\dag h_j - h_i^\dag h_i a_j^\dag a_j - h_j^\dag h_j a_i^\dag a_i - h_i^\dag h_i h_j^\dag h_j \right].
\end{aligned}
\end{equation}
Note that this transformation is exact, provided that one considers initial states with no more than two magnons per site, a subset of the full bosonic Hilbert space.  Of course, all eigenstates of (\ref{tjz}) belong to this subspace and any state in this subspace is confined to it when time evolution is applied.

It is straightforward to check that, up to a constant energy shift, the Green's function of a single hole is exactly the same for the $t$-$J^z$ Hamiltonian $\mathcal H_t + \mathcal H_{J^z} $, given by Eqs.~\eqref{eq:kinetic} and \eqref{hammagint}, and for the interacting boson-fermion model $\mathcal H$ in Eq.~\eqref{model}:

\begin{align}\label{GreensFunctionEquality}
    \mathcal G(\omega + i\delta) &= \left\langle \varnothing \left| h_0 \frac{1}{\omega' + i\delta - \mathcal{H}_t -\mathcal H_{J^z}} h^\dag_0 \right| \varnothing \right\rangle \nonumber \\
    &=\left\langle \varnothing \left| h_0 \frac{1}{\omega + i\delta - \mathcal{H}} h^\dag_0 \right| \varnothing
    \right\rangle,
\end{align}
where $|\varnothing\rangle$ denotes the vacuum state for holes and magnons, $\omega'=\omega-E$. 

To show the validity of the equivalence in Eq. \eqref{GreensFunctionEquality}, let us start with the hole operators $h_i$. For a single hole, the hole-hole interaction term $h_i^\dag h_i h_j^\dag h_j$ vanishes. Moreover, the summation over the hole number operators $h_i^\dag h_i$ yields a constant. Thus these terms can be hidden in the constant energy shift $E$ included in $\omega'$.

For other terms, the equivalence depends on the chosen initial state $\ket{0} = h^{\dag}_0\ket{\varnothing}$. Here, magnons can only be created by the moving hole and there is no separate magnon dynamics. If the initial state contains magnons or if magnons are allowed to move,  equation Eq.~\eqref{GreensFunctionEquality} does not have to be fulfilled. 

States that belong to a linear combination arising from repeated application of the Hamiltonian $\mathcal{H}_t^n \ket{0}$, with $\mathcal H_t$ given by Eq. \eqref{eq:kinetic}, for any $n$ are reachable by the $\mathcal G$ (or $\mathcal{H}_t$) operator, defined in Eq. \eqref{GreensFunctionEquality}, from the initial state $\ket{0}$. On the other hand, any reachable state belongs to $\mathcal{H}_t^n \ket{0}$ for some $n$. Hence, a chain of magnons connects the hole with the site where it was initially created (see Fig.~\ref{fig:1DtJz}). Moreover, there are no other magnons in this state, since magnons can only be created by the hole dynamics. 

Now we need to consider three distinct cases for magnon chains of length $n = 0$, $n = 1$ and $n > 1$. For $n = 0$, the magnon number operator gives $0$ and thus $\mathcal{H} = \mathcal{H}_t + \mathcal{H}_{J^z}$ up to a shift by a constant energy $E$. For $n = 1$, only the hole-magnon interaction and the cost of the creation of a single magnon have to be taken into account (the hole cost can be incorporated into the constant energy shift $E$). It is easy to check that the cost of creating a magnon in a 1D chain is ${J^z}$, and that there will always be only one single hole-magnon interaction (even for longer chains) contributing an energy $-{J^z}/{2}$, so that the total energy contribution of creating the first magnon is ${J^z}/{2}$ and for states with one magnon we again recover $\mathcal{H} = \mathcal{H}_t + \mathcal{H}_J^z$. Creating more magnons in the 1D chain costs no energy, since the energy required to create a magnon is exactly cancelled by the magnon-magnon interaction between neighboring bosons~\cite{Bieniasz2019}. Hence, we show that indeed $\mathcal{H} = \mathcal{H}_t + \mathcal{H}_{J^z}$ for the whole class of reachable states.

Indeed, it is well known that a fermionic quasiparticle appears for any value $J^z>0$ in the $t$-$J^z$ chain with one single hole, agreeing with our results that a quasiparticle is present for any $J\neq 0$ in the fermion-boson model \eqref{model} for $z=2$ with a bosonic point potential acting on the site at which the hole is originally introduced, as given in \eqref{boson-poin-potential} (see Fig.~\ref{fig:1DtJz}) and with the equivalence $J\equiv J^z$. Thus, the spectral function shown in Fig.~\ref{1Dcase} for the limiting value $\tau_0/\tau=\sqrt{2}$ is the same as that for the 1D $t$-$J^z$ with one single hole, where $V/\tau$ is replaced by $J/2t$.

{\it $z>2$ case} For a point potential acting on a Bethe lattice with $z>2$, we can draw a parallel to lightly doped systems with mobile charges. In fact, when doping ions are introduced in an atomic crystal, they affect the crystal structure so that an effective potential is introduced at the point of origin of the hole. This corresponds to $J_n$ given by \eqref{boson-poin-potential}. If we consider the quasi-2D Bethe lattice as a first approximation for 2D systems, we can claim that for lightly doped systems, such that dopants are sparse and distant enough to not affect each other, a quasiparticle will not always stabilize when impurities are introduced in fermionic systems, rather it will depend on the strength of the local potential around the impurity.
A similar situation to this case is also present in the intermediate state of resonant inelastic X-ray scattering (RIXS) on transition metal oxides (TMOs) \cite{Kourtis2012, Ament2011}. Here, introduction of a core hole results in a $3d^{10}$ configuration on the $d$ orbitals, which can be treated as a mobile hole. In turn, this mobile hole is affected by the core-hole potential on the site where it was originally introduced. For TMOs with relatively small magnetic exchange, our model \eqref{model} would then be a good first approximation of the fundamental physics happening at the intermediate state of RIXS, meaning a quasiparticle would stabilize only for large enough values of  core-hole potential.

\section{String-like bosons} \label{sec:StringLike}

In this section, 
 we consider a different shape of the bosonic potential $J_n$ so that 
not one but several bosons have finite energy. 
More specifically, we choose a model in which a finite number of bosons, clustered around the central site of the Bethe lattice, have the same nonzero energy. Without loss of generality, we fix the coordination number $z=3$ in \eqref{model}. This choice corresponds to 
a string-like potential on the Bethe lattice
and to setting the hopping ratio $\tau_0/\tau=\sqrt{3/2}$ in the non-interacting model \eqref{modelham}. 

Thus, we consider the on-site boson energies $J_n$ in Eq.~\eqref{model} as given by
\begin{equation}
    J_n=\begin{cases}
        { J}  & \mbox{if } d_n=0\\
        \frac{ J}{2} & \mbox{if } 0<  d_n < l \\
        0 & \mbox{if } d_n\geq l
    \end{cases}
    ,
    \label{stringpotential-defects}
\end{equation}
which translates in the non-interacting Hamiltonian Eq.~\eqref{modelham} to:
\begin{equation}
    V_{n}=\begin{cases}
        0 & \mbox{if } n=0 \\
        V + \frac{V}{2}(n-1) & \mbox{if } 0<|n|<l\\
        V + \frac{V}{2}(l-1)& \mbox{if } |n|\geq l
        \label{eq:V-string-defect}
    \end{cases}
    .
\end{equation}

{\color{black} For the potential given by Eq.~\eqref{stringpotential-defects} the problem of the corresponding spectral function in general becomes quite complex. Nevertheless, 
the spectral function can be calculated exactly from the continued fraction expansion of the self-energy (Eq.~\ref{self-energy}), as the function $\Gamma(\omega)$ defined in~Eq.~\eqref{TailExpansion} can be easily solved,
reducing the problem of calculating the infinite continued fraction to a finite one, see discussion after Eq.~\eqref{self-energy} and Appendix \ref{App:A} 
for details. On the other hand, it is not possible to provide a general analytic form for the quasiparticle spectral weight and energy -- except once $l=1,2$ or 
$l \rightarrow \infty$ in Eq.~\eqref{stringpotential-defects}, see Appendix \ref{App:B} for details.}
\subsection{Solution}

\begin{figure}[t!]
\includegraphics[width=\columnwidth]{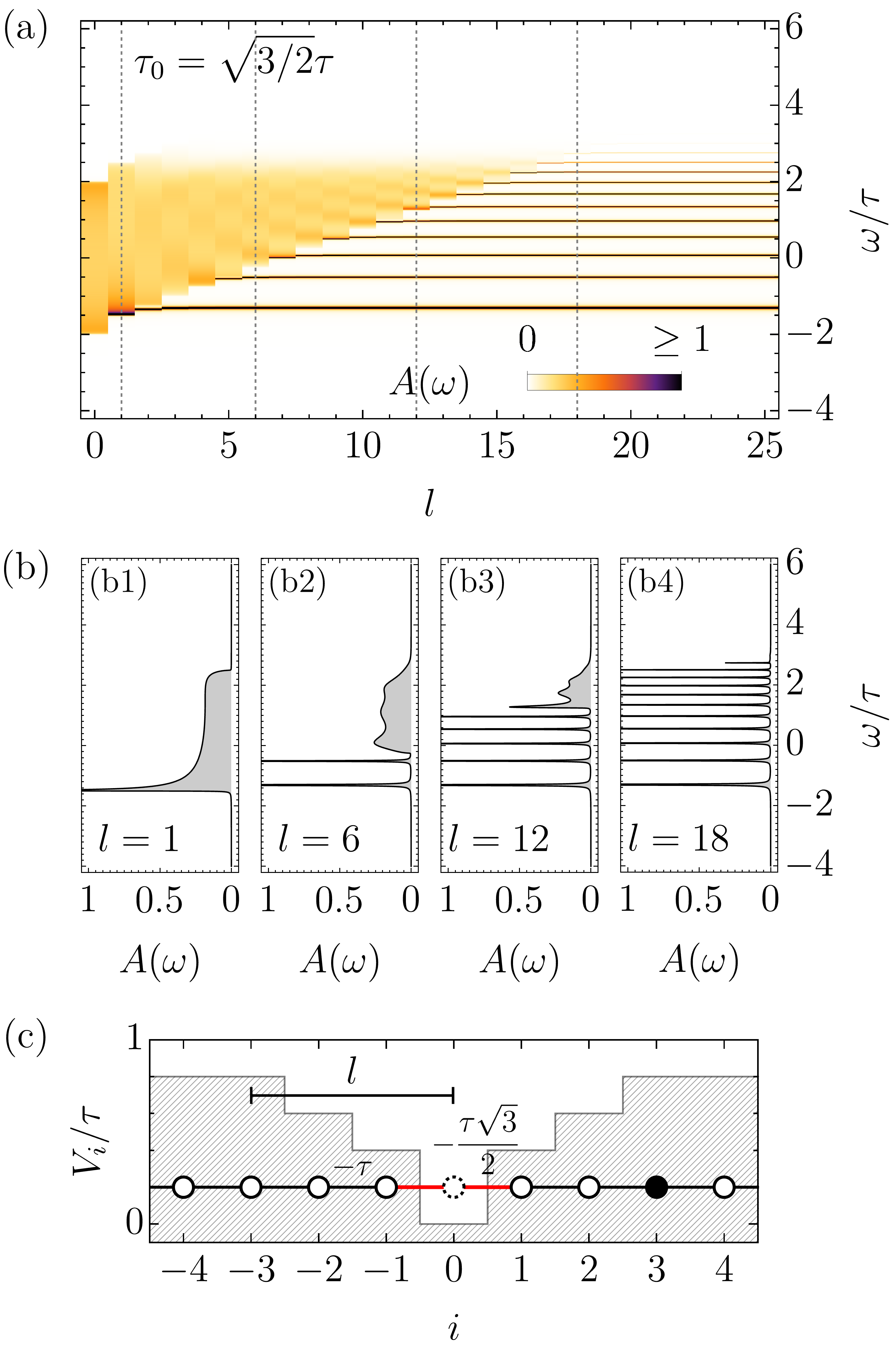}
    \caption{Spectral function $A(\omega)$ for a string-like potential [cf. \eqref{eq:V-string-defect}] and varying half-width of the potential well $l$ for ${\tau_0}/{\tau} = \sqrt{3/2}$ and the (right panel) potential parameter $V=0.5\tau$. A transition from a solution with no quasiparticles to one with multiple quasiparticles is observed.
    (b) Vertical cuts of panel (a) at different values of the potential half-width $l$. (b1) $l=1$, no discrete state peak is present, (b2) $l=6$ two discrete state peaks are present in the spectrum as well as a continuum, (b3) $l=12$ several discrete peaks are visible as well as a higher energy continuum of states, (b4) $l=18$ several discrete peaks make up the spectrum, similar to the known ladder-spectrum.
    (c) Schematic of the model in a string potential with defects [cf. \eqref{eq:V-string-defect}]. The different hopping around the origin is drawn in red, the gray area shows the altered string potential, the black filled circle represents the moving fermion and the dashed circle its starting position, we explicitly shown the half-width of the potential~$l$.}\label{stepconst}
\end{figure}


We plot the spectral function for the string-like potential \eqref{eq:V-string-defect} and $\tau_0/\tau=\sqrt{3/2}$ in Fig.~\ref{stepconst}. 
As
the relevant parameter in this case 
we use
the half-width of the potential $l$, while the potential parameter is 
set to be $V=0.5\tau$.


To better show the progression of the spectral function $A(\omega)$ as a function of $l$, we present $A(\omega)$ for four values of $l$ in Fig.~\ref{stepconst}(b). For $l=1$ (Fig.~\ref{stepconst}(b1)), no discrete state is present. Note that this is the same situation shown in the vertical cut of Fig.~\ref{pointpotential}(a), since $l=1$ corresponds to a point potential. Now we focus on increasing $l$. At $l=6$, already two distinct discrete states are present, together with a relatively large continuum of states (cf. Fig.~\ref{stepconst}(b2)). When $l=12$, the spectral function shows several consecutive peaks, but a continuum persists for larger energies $\omega/\tau$ (cf. Fig.~\ref{stepconst}(b3)). Finally, for $l=18$, we see a spectrum composed only of consecutive sharp peaks, i.e., discrete states, similar to a ladder spectrum. This situation keeps on as $l$ goes to infinity. It is then natural to consider the special limit of a full string potential covering the whole lattice. 


\subsection{Special Limit}

\begin{figure}
    \includegraphics[width=\columnwidth]{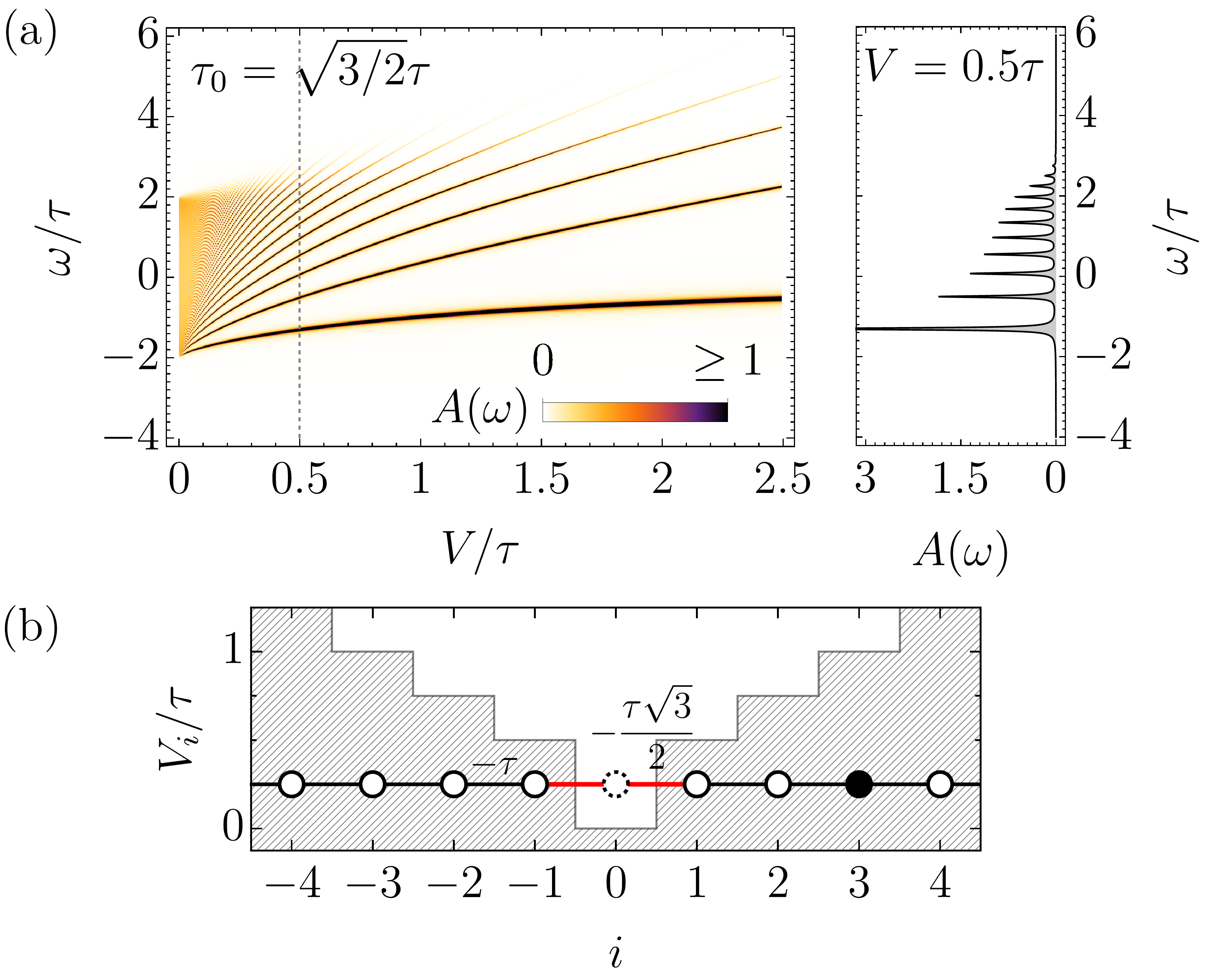}
    \caption{Spectral function $A(\omega)$ in a string potential [cf. \eqref{eq:V-string}] for $\tau_0/\tau=\sqrt{3/2}$. The right panel shows the spectral function $A(\omega)$ at $V=0.5\tau$: the spectral function takes the shape of the so-called ladder spectrum
    and contains only discrete (i.e. quasiparticle-like) peaks. 
    (b) Schematic of the model at the vertical cut drawn in (a). The different hopping around the origin is drawn in red, the gray area shows the string potential, the black filled circle represents the moving fermion and the dashed circle its starting position. Note that the string potential completely covers the infinite chain.}
    \label{fig:BetheSpectralFunction}
\end{figure}  

We now consider the limit $\l \to \infty$. This results in a finite on-site bosonic energy $J_n$  for $every$ site $n$ on the Bethe lattice
in Eq.~\eqref{model}:
\begin{equation}
    J_n=\begin{cases}
        J & \mbox{if } d_n=0 \\
        \frac{J}{2} & \mbox{if } d_n> 0,
    \end{cases}
    \label{stringpotential}
\end{equation}
i.e., {\it all} bosons are massive.
This situation corresponds to a potential $V_{n}$ in the non-interacting Hamiltonian Eq.~\eqref{modelham}
having a {\it perfect} string (i.e. discrete linear) character:
\begin{equation}
    V_{n}=\begin{cases}
        0 & \mbox{if } n=0 \\
        V+\frac{V}{2}(n-1) & \mbox{if } n\neq 0.
        \label{eq:V-string}
    \end{cases}
\end{equation}

In Fig.~\ref{fig:BetheSpectralFunction}, we show the spectral function for such a model as a function of the potential parameter $V/\tau$.
There is an important difference in the physics depicted in Fig.~\ref{stepconst} and Fig.~\ref{fig:BetheSpectralFunction}.  In the first case, we plot the spectral function in terms of the half-width of the potential $l$, i.e., in terms of how many sites are actually affected by it and we fix the potential parameter to be $V=0.5\tau$. In the second case, we show the spectral function evolution as a function of the potential energy parameter $V/\tau$, similarly to what done previosuly for the case of a point potential.
Indeed, we observe that
this case is very different from the previously considered case of a point potential, since there exist at least one discrete energy state for any value of $V$. However, as the potential increases, we see that more discrete states appear in the system, in contrast to the point potential case were only one discrete state would stabilize.

The stark contrast to the case of impurity-like bosons [cf. Sec.~\ref{sec:PointPotential}] is underlined
when we fix the value of $V$ and plot $A(\omega)$: a structure with several consecutive delta-like peaks appear, 
 [right panel of Fig.~\ref{fig:BetheSpectralFunction}(a)]. This structure is the  well-known
ladder-spectrum of~\cite{Bulaevskii1968, Kane1989}.
In the language of the fermion-boson model this means that for any value of the finite bosonic energy 
$J$ the spectral function of model~\eqref{modelham} with \eqref{stringpotential} {\it solely} consists of quasiparticles, cf.~\cite{Bulaevskii1968, Kane1989}.

Note that at least one discrete state emerges for any value of $V$, in contrast with the finite string potential case of Fig.~\ref{stepconst} for which for any fixed value of V, we have a continuum until a critical value $l^*$ is reached, meaning the discrete state appears only when the potential `covers' enough sites around the origin, where the fermion is introduced. This number depends on the parameter $V$ as well as the hopping ratio $\tau_0/\tau$.  In terms of the interacting fermion-boson model, this translates to the possibility of having quasiparticle decay in case only some bosons around the origin of the Bethe lattice are massive -- with the rest being massless. Nonetheless, to reach the ladder spectrum form when $l$ is finite, the potential needs to cover a relatively large number of lattice sites. Hence, it is rather easy to destabilise the ladder spectrum and recover some kind of state continuum.

In the language of the interacting fermion-boson model this implies that, whereas in the case presented in Fig.~\ref{fig:BetheSpectralFunction} we are tuning the strength of the boson-fermion coupling $t/J$ (strictly speaking $J/t$ in Fig.~\ref{fig:BetheSpectralFunction}), in the case of  Fig.~\ref{stepconst}
we are tuning the number of massless bosons $\propto l$ present in our system.

\subsection{Physical realization} \label{sec:physrealstringlike}
\begin{figure}[t!]
    \centering
    \includegraphics[width=0.8\columnwidth]{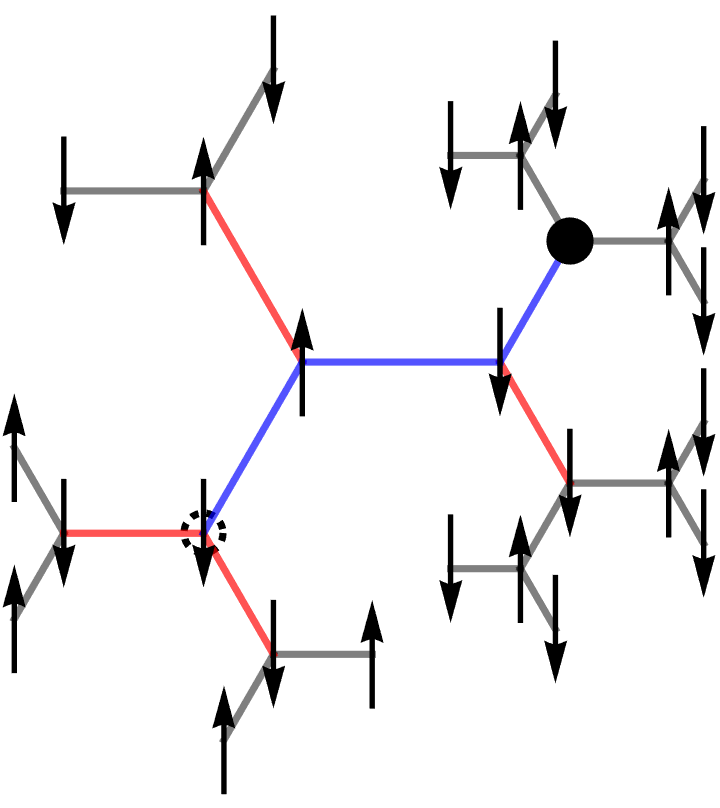}
    \caption{Dynamics of a single hole on a Bethe lattice with $z=3$. The black solid circle represents the hole in the current state, the dashed circle is the hole at its initial creation site. The cartoon shows the system after few hole hoppings have occurred.}
    \label{fig:Bethe}
\end{figure}

{\it Perfect string potential} The spectrum in Fig.~\ref{fig:BetheSpectralFunction} immediately brings to mind the one of the $t$-$J^z$ model on a Bethe lattice with $z>2$~\cite{Bulaevskii1968, Kane1989}. 
While the relation between the fermion-boson model in Eq.~\eqref{model} and the $t$-$J^z$ model on a Bethe lattice was studied earlier in \cite{Chernyshev1999, Bieniasz2019, Wrzosek2021}, for completeness let us describe it in detail here.
It is straightforward to generalize Eq.~\eqref{tjz} to any $z>2$. The subsequent consideration also apply, especially the equivalence of the Green's functions shown in Eq.~\eqref{GreensFunctionEquality}, with the shape of the bosonic potential as given in \eqref{stringpotential}. More generally, equality \eqref{GreensFunctionEquality} holds true for any coordination number $z$ and the choice of $J_n$
\begin{equation}
    \label{eq:Ji}
\begin{split}
    J_0 &= \frac{J^z}{2}(z-1), \\
    J_{n\neq 0} &= \frac{J^z}{2}(z-2)d_n
\end{split}
\end{equation}
and the site $n=0$ is the site at which the hole is originally introduced in our lattice (see Fig.~\ref{fig:Bethe}). While the consideration about the energy shift due to the presence of a hole stay the same, generalizing the energy equivalence for the cases when  magnons are present in \eqref{GreensFunctionEquality} requires some attention, since now we are considering a Bethe lattice with $z>2$ and there are more neighboring sites. However, the analysis of states reachable through $\mathcal H_t$ still holds true, so that the motion of the hole on a Bethe lattice still results in a chain of magnons connecting the hole to the site where it was initially created. Again, there are no other magnons in the system, since only the hole dynamics can result in magnon creation. Thus, the magnons and the hole form a 1D-like chain on the Bethe lattice branch.

It is again necessary to consider three distinct cases of magnon chains of length $m=0$, $m=1$ and $m>1$. For the $m=0$ case, the same conclusions as in the $z=2$ case apply, so that $\mathcal H = \mathcal H_t + \mathcal H_{J^z}$. However, we now generalize the energy cost of creating a magnon to include any value of $z$. This cost is  found to be ${zJ^z}/{2}$. Again, there will always be only one single hole-magnon interaction (even for longer effective chains) contributing an energy $-{J^z}/{2}$. Thus, setting the cost of the first magnon to $J_0 = ({J^z}/{2})(z-1)$, we see that also for states with one magnon we have $\mathcal{H} = \mathcal{H}_t + \mathcal{H}_J^z$. For $m > 1$ we need to take into account the magnon-magnon interaction. Since magnons form a 1D-like chain, then number of such interactions will be $m - 1$ and each one of them will contribute an energy $-J^z$, cf.~\cite{Wrzosek2021}. This lowers the energy of every magnon beyond the first one, so that choosing $J_{n>0} = ({J^z}/{2})(z-2)d_n$ we show that indeed $\mathcal{H} = \mathcal{H}_t + \mathcal{H}_{J^z}$ for the whole class of reachable states.

Note that, when $z=3$, we recover \eqref{stringpotential}. 
We can conclude that Fig.~\ref{fig:BetheSpectralFunction} exactly shows the spectral function for a single hole moving on a Bethe lattice with $z=3$ hosting the $t$-$J^z$ model, depicted in Fig.~\ref{fig:Bethe}.

{\it String-potential with defects} 
The most accurate realisation of a fermion-boson system with bosons subject to a string potential with defects as considered by Eq.~\eqref{stringpotential-defects}
is as follows. Let us assume that we have a system which consists of two kinds of subsystems: an Ising ferromagnet and an Ising antiferromagnet.
Next, we put these two subsystems on a Bethe lattice 
in such a way that the antiferromagnet surrounds
the origin of the Bethe lattice and the ferromagnet starts $l$ sites away from the origin. Finally, once we probe such a system with the local fermion Green's function, by putting a mobile hole into the origin of the Bethe lattice, then such a problem is described exactly by model~\eqref{model}-\eqref{spectral-function} with the potential~\eqref{stringpotential-defects}. This is because: (i) the hole in the antiferromagnetic subsystem is described by \eqref{model} with the constant bosonic energies $J_n$ given by Eq.~\eqref{stringpotential} -- see discussion immediately above, (ii) the hole in the ferromagnetic subsystem can freely move without introducing spin flips -- which in the language of model \eqref{model} means that the hole excites magnons with zero-energy, i.e. $J_n=0$ for $d_n>l$ on a Bethe lattice.

\begin{figure}[t]
    \centering
    \includegraphics{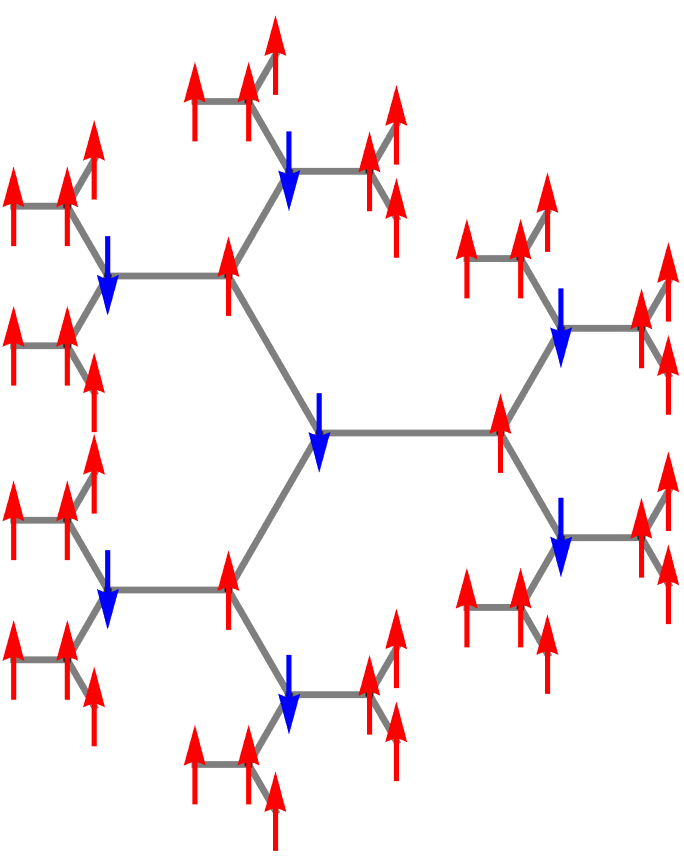}
    \caption{An example of an antiferromagnetic domain of size $d=3$ embedded in a ferromagnet on a Bethe lattice with $z=3$.}
    \label{fig:AFM-in-FM}
\end{figure}

The aforementioned ferromagnetic-antiferromagnetic
system can be realised by considering antiferromagnetic domains
of size $\propto l$ immersed in a ferromagnet, as sketched in Fig.~\ref{fig:AFM-in-FM}. 
Such a situation, to some extent, may occur in the overdoped cuprates. In this case the tendency to ferromagnetism upon doping~\cite{Kopp2007, Jia2014, Santoso2017, Ong2022} and diminishing antiferromagnetic correlations with doping~\cite{Lee2006, Zhang2022} may, in the first-order quantum phase transition scenario~\cite{Kopp2007} or phase separation scenario~\cite{Battisti2016}, locally lead to the onset of antiferromagnetic domains within the ferromagnetic background. Naturally, there are several differences between the cuprate models (such as doped $t$--$J$ or Hubbard) and the model considered here -- the main ones concern the Heisenberg (and not Ising) spin exchange, the finite number of holes (and not just one single hole), and the 2D square lattice (and not Bethe lattice with $z>2$). Nevertheless, we believe that these differences are not important enough to fully hinder the applicability of the current study to the cuprate problem. Thus, the interesting insight gained from the problem under study here is that, if the antiferromagnetic domains are very small and the system is `almost' ferromagnetic, a fermion inserted into the antiferromagnetic subsystem may not form a well-defined quasiparticle
(especially if the antiferromagnetic exchange is small). On the other hand, 
if the antiferromagnetic domains are relatively large (compared to the ferromagnetic ones), the quasiparticle may be stabilised relatively easily. This counterintuitive conclusion can be tested experimentally by STM studies of the overdoped cuprates.

A somewhat similar situation can occur in correlated systems with orbital degrees of freedom and incommensurate filling -- such as doped manganites or vanadates~\cite{Dagotto20011, Miyasaka2000, Fujioka2005, Fujioka2008, Avella2019}. In this case hole doping induces transition from the alternating orbital to the ferro-orbital state. Thus one may expect that upon moderate doping, and assuming the onset phase separation, domains with alternating orbital order would become surrounded by the ferro-orbital ordered state. Then, depending on the relative size of such alternating orbital domains, the quasiparticle decay may (small domains) or may not (large domains) happen . Note that in the orbital case we are closer to model~Eq.~\eqref{model} with \eqref{stringpotential-defects} than in the cuprate case described above, since the orbital degrees of freedom interact in correlated systems in an almost Ising manner~\cite{Kugel1982}. 

One can also think of other realisations of such a string potential with defects. This may for instance occur once a hole is introduced in the antiferromagnetic side of a ferromagnetic-antiferromagnetic interface (or alternatively on the alternating orbital side of a
ferro-orbital-alternating orbital interface). Here, however, a detailed investigation is needed, for the topology of such a problem is quite different:
in this case one of the subsystems is not entirely surrounded by the other one. Nevertheless, the intuition gained by the current study suggests that also in this case the quasiparticle could decay if the size of the ferromagnetic (or ferro-orbital) subsystem is significantly larger than the antiferromagnetic (alternating orbital) one.
Finally, one also speculates that there may exist also other physical systems with fermions coupled to zero-energy bosons in one subsystem and to bosonic excitations with finite energy in another. This might be the case in a rather exotic 
situation in which one of the subsystems contains condensed bosons in real space and the other one is a `normal' state with bosonic excitations costing finite energy.

\section{Discussion \& Conclusions} \label{sec:Discussion}

In this work, we have considered an interacting fermion-boson model with {\color{black} bond-type polaronic coupling between fermions and local bosons}. We have solved it analytically by mapping it to an impurity-like non-interacting chain. The fermion-boson coupling introduced in model \eqref{model} can take several forms, but we have restricted ourselves to two important cases:
the impurity-like and string-like bosons
(future studies of other cases are encouraged).
We have shown how in these two cases it is possible to destabilize the fermionic quasiparticle either by tuning the relative strength of the fermion-boson coupling or by increasing the number of zero-energy bosons:

When considering {\it impurity-like bosons} in Sec.~\ref{sec:PointPotential}, one can tune the relative strength of the fermion-boson coupling $t/J$ such that no quasiparticle appears in the system for $J\leq J^*$ (for fixed $t=1$), with $J^*$ depending on the Bethe lattice coordination number $z$.
On the other hand, for $J > J^*$ the quasiparticle is stabilised.
Overall, the critical value $J^*$ increases with the coordination number $z$ -- but $z=2$ is a limiting case with $J^*=0$, i.e., we always obtain a quasiparticle solution (see below).
Note that for the impurity-like bosons the number of zero-energy bosons cannot be tuned (it equals the number of lattice sites minus one, for the bosons have a hard core).
Thus, it is interesting to observe that, despite the coupling to an overall huge number of massless bosons, a fermionic quasiparticle is still  stable once $J > J^*$.

For {\it string-like bosons} as in Sec.~\ref{sec:StringLike}, the situation is more complex, for we can tune here both the number of zero-energy bosons as well as the coupling between fermions and bosons.
First, just as for the impurity-like bosons, an increase in the relative fermion-boson coupling strength $t/J$ can destabilize the quasiparticle. 
Second, the stability of a quasiparticle depends on the number of bosons with zero-energy.
On the one hand, if there are no zero-energy bosons, the hole is affected by a discrete linear  potential (string potential), since each created boson costs energy. 
In this case all of the eigenstates of the system
are of quasiparticle-type and we obtain the so-called ladder spectrum (see below). On the other hand,
the ladder spectrum of the perfect string-potential is rather fragile, since a finite number of bosons with zero-energy will result in a decrease in the number of quasiparticles, the emergence of an energy continuum and, for the critical number of these bosons, the onset of a completely incoherent spectrum. The latter takes place for a relatively large number of massless bosons 
and strong fermion-boson coupling $t/J$.

Lastly, we have mapped the fermionic Green's function of the fermion-boson model
in {\it specific} limits to that of the single hole in the $t$-$J^z$ model. We have shown that: (i) the fermion-boson model with impurity-like bosons and the coordination number $z=2$ corresponds to the 1D $t$--$J^z$ model; (ii) the fermion-boson model with string-like bosons subject to a perfect string potential and with coordination number $z>2$
corresponds to the quasi-2D (Bethe lattice) $t$--$J^z$ model. Note that these two particular limits carry a quasiparticle solution for any finite value of the model parameters, as discussed above and as  well-known from the extensive $t$--$J^z$ model literature, cf.~\cite{Bulaevskii1968, Brinkman1970, Kane1989, 
Starykh1996,  Sorella1998, 
Chernyshev1999,
Smakov2007,
Smakov2007b,
Maka2014,
Grusdt2018, Bieniasz2019, Wrzosek2021}.

Finally, it is possible to slightly modify these particular limits of the fermion-boson model to destabilise the quasiparticle solution
(thus, e.g. in 1D one can scale the first hopping of the hole in the $t$--$J^z$ model; in 2D one can add an Ising ferromagnetic interface next to the Ising antiferromagnet).
Nevertheless, it is a stunning observation of this work that the parameter range
for which the quasiparticle decay happens is small when compared to the range for which it is stable -- and this is despite the ubiquitous presence of zero-energy bosons in the system. 


%

Examples of boson-fermion systems in which the
quasiparticle collapse may happen include systems with a low concentration of impurities and mobile fermions (for instance created in the intermediate state of RIXS experiment). Another example concerns fermions
introduced to antiferromagnetic domains immersed in the ferromagnetic background or alternating orbital domains immersed in the ferro-orbital background. Rather counterintuitively, the quasiparticle extinction in these cases
is more likely once the ferro-ordered state
dominates.
Such a situation might take place in overdoped cuprates -- or other doped transition metal oxides with orbital degrees of freedom (e.g. manganites or vanadates). Further
experimental and theoretical studies are needed to verify the latter proposal.

\section*{Acknowledgements}

This work was supported by Narodowe Centrum Nauki (NCN, Poland) under Project Nos. 2016/22/E/ST3/00560 and 2021/40/C/ST3/00177.

For the purpose of Open Access, the author has applied a CC-BY public copyright licence to any
Author Accepted Manuscript (AAM) version arising from this submission.

The code to reproduce the data and figures presented in this manuscript is available at Ref. \cite{zenodo}.

{\color{black}
\begin{appendix}

\section{Remarks on the the exactness of the Greens function calculations} \label{App:A}

One of the central observables presented in this work is the single particle spectral function $A(\omega)$ defined in Eq.~\eqref{eq:SpectralFunction} in the main text. Although the system we investigate is infinite, in all the presented cases we were able to present the numerically exact results. Here we explain how this is achieved. 

Let us assume a potential $V_i$ acting on a particle in Hamiltonian~\eqref{modelham} such that for $i \geq l$ the potential stays constant (i.e. $V_{l} = V_{l+k}$ for $k=1,2,3,...$) but it is otherwise arbitrary, yet finite.
Note that this assumption allows for studying 
all but one case discussed in the main text of the paper---i.e. both
the case of a point potential as well as more sophisticated step potentials discussed in the main text of this paper
(whereas the one case with $l \rightarrow \infty$ is discussed below).
Provided that $l$ is reasonably small (up to several millions), the continued fraction expansion of the self-energy provided in Eq.~\eqref{self-energy} can be evaluated directly. This is thanks to the fact that it terminates with a function $\Gamma(\omega)$ [Eq.~\eqref{TailExpansion}], which can be solved analytically with standard methods yielding two solutions,
\begin{equation} \label{eq:a1}
    \Gamma(\omega) = \frac{\omega - V_l}{2} \pm \frac{1}{2} \sqrt{(\omega - V_l)^2 - 4\tau^2}.
\end{equation}
When calculating the spectral function with a finite broadening, the sign choice $\pm$ should follow the sign of $V_l - \operatorname{Re} \omega$. This way only a finite number of terms in the continued fraction have to be calculated.

On the other hand, for $l \to \infty$, i.e. in the case of the ``ideal'' step potential presented in the paper, the solution for the Greens function can be expressed as a ratio of two Bessel functions of the first kind~\cite{Wrzosek2021,Bieniasz2019}.

\section{Remarks on the existence of analytical solutions to the quasiparticle energy and weight}
\label{App:B}

In general, the fact that $\Gamma(\omega)$ can be calculated exactly [see \eqref{eq:a1}] allows us to express the energies of the quasiparticle states of the investigated model in terms of roots of a polynomial of a finite degree (instead of the infinite one). This polynomial is given by,
\begin{equation}
    Q(\omega) = \left(2P_l(\omega) - (\omega - V_l)P_{l-1}(\omega) \right)^2 - (\omega - V_l)^2 - \tau^2,
\end{equation}
where $P_k$ can be evaluated according to the following recurrence relation,
\begin{equation}
    \begin{pmatrix}
        P_k(\omega) \\
        P_{k-1}(\omega)
    \end{pmatrix} = 
    \begin{pmatrix}
        \omega - V_{k-1} & -\tau^2 \\
        1 & 0
    \end{pmatrix}
    \begin{pmatrix}
        P_{k-1}(\omega) \\
        P_{k-2}(\omega)
    \end{pmatrix},
\end{equation}
with
\begin{equation}
    P_1(\omega) = \omega - V_0 \quad \mathrm{and} \quad P_0(\omega) = \frac{\tau_0^2}{\tau^2}.
\end{equation}
Since $Q(\omega)$ is a polynomial of degree $2l$, the case of a point potential ($l=1$) and 2-step potential ($l=2$) can be solved analytically. For polynomials of degree higher than 4 the algebraic (finite) solution (in terms of radicals) does not exist in general. In this case, the quasiparticle energy and weight typically have to be evaluated numerically.

On the other hand, for the infinite step potential ($l \rightarrow \infty$) the quasiparticle weight can be expressed through an infinite series of Bessel functions $\mathcal{J}_{\alpha}(x)$,
\begin{equation}
    a_\mathrm{QP} = \frac{1}{1 + \frac{\tau_0^2}{\tau^2}\sum\limits_{n=1}^{\infty} \left(\frac{\mathcal{J}_{1-\frac{2\varepsilon_\textsc{gs}}{V} + n}\left(\frac{4\tau}{V}\right)}
{\mathcal{J}_{1-\frac{2\varepsilon_\textsc{gs}}{V}}
\left(\frac{4\tau}{V}\right)}\right)^2}.
\end{equation}
Importantly the terms under the sum in the denominator decay super-exponentially with $n$ leading to an extremely fast convergence of the above equation~\cite{Bieniasz2019}. The quasi-particle energy $\varepsilon_\textsc{gs}$ may be evaluated numerically from the relation,
\begin{equation}
    \varepsilon_\textsc{gs} - V_0 = \Sigma(\varepsilon_\textsc{gs}) = -\frac{\tau_0^2}{\tau}\frac{\mathcal{J}_{2-\frac{2\varepsilon_\textsc{gs}}{V}}\left(\frac{4\tau}{V}\right)}
{\mathcal{J}_{1-\frac{2\varepsilon_\textsc{gs}}{V}}
\left(\frac{4\tau}{V}\right)}.
\end{equation}
    
\end{appendix}
}

\end{document}